\documentclass[12pt] {article}
\usepackage[affil-it]{authblk}
\usepackage{amsmath}
\usepackage{physics}
\usepackage{epsfig}
\usepackage{amsfonts}
\usepackage{amssymb}
\usepackage{geometry}
\usepackage{graphicx}
\usepackage{braket}
\allowdisplaybreaks
\newcommand {\bp}{\begin{pmatrix}}
\newcommand {\ep}{\end{pmatrix}}
\newcommand{\be}{\begin{equation}} \newcommand{\ee}{\end{equation}}
\newcommand{\bea}{\begin{eqnarray}}\newcommand{\eea}{\end{eqnarray}}

\begin{document}

\title{Entanglement Induced  by  Noncommutativity:\\  Anisotropic Harmonic Oscillator in Noncommutative space}

\author{Abhishek Muhuri%
  \thanks{Electronic address: \texttt{abhishekmuhuri96@gmail.com}}}
\affil{Department of Physical Sciences, Indian Institute of Science Education and Research, Kolkata, Mohanpur 741246, India}
\author{Debdeep Sinha%
  \thanks{Electronic address: \texttt{sinha.debdeep@gmail.com}}}
\affil{Physics and Applied Mathematics Unit, Indian Statistical Institute\\
203 B.T.Road, Kolkata 700108, India}
\author{Subir Ghosh%
  \thanks{Electronic address: \texttt{subirghosh20@gmail.com}}}
\affil{Physics and Applied Mathematics Unit, Indian Statistical Institute\\
203 B.T.Road, Kolkata 700108, India}

\maketitle

\abstract {Quantum entanglement, induced by spatial noncommutativity, is investigated for an anisotropic harmonic oscillator. Exact solutions for the system are obtained after the model is re-expressed in terms of canonical variables, by performing a particular Bopp's shift to the noncommuting degrees of freedom. Employing Simon's separability criterion, we find that the states of the system are entangled provided a unique function of the (mass and frequency) parameters obeys an  inequality. Entanglement of Formation for this system is also computed and its relation to the degree of anisotropy is discussed. It is worth mentioning that, even in a noncommutative space, entanglement is generated only if the harmonic oscillator is anisotropic.   Interestingly, the Entanglement of Formation  saturates for  higher values of the deformation parameter $\theta$, that quantifies  spatial noncommutativity. }

\newpage

\section{Introduction}

The  celebrated
ideas  of quantum
computers, quantum cryptography, and quantum teleportation have emerged from quantum information science that deals with manipulation of individual quanta of information.  Entanglement has assumed the role of a key resource in quantum communication and computation and it is exploited as  protocols  to perform tasks that are supposedly intractable with classical information processing. Entanglement lies at the heart of most of the counter intuitive features of  quantum mechanics, such as  non-local correlations between space like separated entangled partners (for a review see \cite{enrev}) and related phenomena. In spite of Einstein's worry about the "spooky action at a distance", it has been experimentally confirmed \cite{enexp1,enexp2} that quantum physics predicts correlations that violate Bell's inequality. However, as it turns out, quantum entanglement has found  remarkable physical applications  in the fields like quantum computation, quantum optics, condensed matter physics and many more\cite{mbp}\cite{bs}\cite{nl}. Connections between quantum gravity and quantum information theory in the $AdS/CFT$ correspondence, in particular relation between geometrical structure of the dual spacetime and entanglement structure of the conformal  field theory have emerged in recent times (for review see \cite{mvr}). Another topical and exciting area is entanglement entropy \cite{enent} and its application in the issue of black hole entropy (for a review see \cite{re}).

In the present article we will consider a novel form of entanglement generation - Noncommutativity induced entanglement. Earlier works in this area can be found in \cite{n15}, \cite{n1}, \cite{n6}. In fact the work by Bastos et. al. \cite{n6} is very significant in this context and we will return to its relevance in connection to the present work in section 6,  Conclusion. Noncommutative (NC) extension of quantum mechanics and quantum field theory have generated a large amount of interest in diverse physics communities (for reviews see \cite{ncrev}). Historically NC extension was first introduced by Snyder \cite{sny} to ameliorate short distance singularities in quantum field theories but it was not successful in this context (although for other reasons Snyder's form of Noncommutativity has also become of interest in recent times (\cite{ncrev1})). Another form of NC extension induces a minimal length and is exploited in phenomenological model building for quantum gravity \cite{kemp, am} and in addressing certain conceptual inconsistencies in black hole physics \cite{dfr}. The more recent wave of interest was created by the work of Seiberg and Witten \cite{sw} who revealed that in certain low energy limits, the theory of open strings attached to $D$-Branes, can be described as an  effective field theory residing in  an NC space. Interestingly, this form NC extension actually mimics the well known Landau problem of planar charged particle dynamics in a strong uniform background magnetic field in normal direction. This last form of noncommutativity will be directly relevant to our work. Some of the earliest applications of noncommutative geometry in explicit quantum mechanical problems appeared in \cite{n8}, \cite{n11}, \cite{n16}. Formal
aspects of noncommutative quantum mechanics appeared in \cite{n3}, \cite{n4}, \cite{n12}. In the context of gravitational physics, neutrons in the presence of an external 
gravitational field with noncommutativity were studied in \cite{n7}. Novel effects, induced by noncommutativity, were considered in    quantum cosmology \cite{n5}, \cite{n13} and in  black hole physics \cite{n2}.  Aspects of  noncommutative geometry were discussed in \cite{n10} in the context of   string theory  and in \cite{n9, dv, s1, s2,s3,s4,s5} in quantum field
theory.

Let us come to our main subject matter: entanglement and in particular Gaussian Entanglement (GE). Entanglement is manifested through 
nonlocal correlations between two (or more) quantum states such that it is impossible to describe  the states individually.  Operationally this trait implies the presence of global states of a composite system that are not expressible as  products of  states of the individual subsystems. 
Quantum information can appear in discrete and
continuous forms. Initial  studies on
entanglement were concentrated  on finite-dimensional quantum systems of which the simplest and most well-known example of discrete quantum
information is the quantum bit or ‘‘qubit,’’ a quantum system
with two distinguishable states. However, quantum
information can also be encoded in  a continuous variable quantum system where it has an infinite dimensional Hilbert space spanned
by observables with continuous eigenspectra. It has been realized that the latter approach, 
where information carriers are continuous variables (rather than discrete qubits)  constitutes an extremely powerful alternative framework for  quantum information processing (for reviews see \cite{gauss1,gauss2}).   Continuous
variable systems have led to the advances in resource states preparation, fault tolerant
quantum computation as well as   analysis of topological
order, cryptography and machine learning in the theoretical domain.

Gaussian states bear a special significance as continuous variable states: \\
(i) Even though  being infinite-dimensional systems  they are characterized by a  finite
number of parameters.\\
(ii) A simple algebraic
formalism can be used to manipulate them analytically.\\
(iii) Any quantum information process depends on  combination of Gaussian states,
Gaussian operations, and Gaussian measurements. \\
(iv) The physical states and algebraic operations can be realized and controlled in the
laboratory utilizing  standard techniques of quantum optics such as beam splitters, phase shifters, squeezers and other efficient detection systems. 

Gaussian states can be formed in a variety of physical systems that include light field modes \cite{lf,lf1,lf2}, cold atoms \cite{vj}, excitons in photonic
cavities \cite{gly}. GE between two Gaussian modes are readily  generated, for example, in case of two output beams of a parametric 
down converter  sent through optical fibers \cite{op} or in atomic ensembles interacting with light \cite{al}. Apart from that, Gaussian states find applications in realizations of quantum key distributions \cite{fgg},
teleportation \cite{tcz} and electromagnetically induced transparency \cite{dak}. Technique has also been developed to quantify GE theoretically \cite{rr,rs,lm, ggb,spd}. In this perspective, it is indeed worthwhile  to study alternative ways of generating  GE. The present work discusses a  novel possibility: inducing entanglement by introduction of an additional structure 
in phase-space manifold, i.e, the {\it spatial noncommutativity}.

In the present paper we investigate GE in an anisotropic Harmonic Oscillator (HO) in NC space. Let us briefly outline the flowchart of our work. We start by writing down the Hamiltonian for the two-dimensional anisotropic HO in NC coordinates and momenta. Subsequently we exploit Bopp's shift method to rewrite the Hamiltonian in canonical variables where it is easy to see that indeed the noncommutativity is capable of generating GE. For computation of the entanglement entropy, we exploit a simple relation, derived by Rendell and  Rajagopal \cite{rr}. This requires  exact solution for the NC anisotropic HO model which we derive following the framework used in Ref. \cite{ql}.   Finally we compute the GE for our model. Interestingly, under certain approximations, our model can be identified with  a planar HO in external magnetic field, perpendicular to the plane where the noncommutativity of the space generates an effective magnetic field.

 Previously in the literature, separability criterion of oscillator systems has been discussed in terms of the covariance 
matrix \cite{axb}. However in this paper, we take a step ahead and provide a quantitative measure for such entanglement as well as analyze
 Simon's separability condition for our system. It turns out that the solutions of the anisotropic HO in NC space are not separable, thus implying 
that the system is entangled. It is also observed that this entanglement has an upper-bound in high deformation region and it increases with the deformation 
parameter asymptotically to this value.

Noncommutativity induced entanglement entropy for an {\it isotropic} two dimensional HO was studied in Ref. \cite{bl}. There are nontrivial differences in the formalism adopted: In Ref. \cite{bl} the authors have worked  in a  Wigner function framework with NC degrees of freedom whereas we have utilized the well established Bopp's shift method to convert the NC variables to canonical ones and have directly computed the von Neumann entanglement entropy for a most general form of {\it anisotropic} two dimensional HO. We will compare and contrast our results later.

The sections are arranged in the following way: In Section 2, we shall present a brief review of the work done so far on the entanglement properties 
of HO and state the main results that are useful in the present study. Mathematical details are provided in  Appendix A. 
 In Section 3, we shall introduce our model and discuss in an intuitive manner why we expect entanglement
to be present in the system. Further, we shall also obtain the exact solutions for the anisotropic HO in NC space.
The  rationalization procedure for the Hamiltonian is interesting but quite involved. The details can be found in Appendix B. Section 4 is devoted to the study of GE properties of our model . It contains our major result comprising of  the measure of GE. We discuss our results graphically. In Section 5 we make a summary of the work and discuss our results.

\section{Entanglement properties of pure two-mode Gaussian state}

In this section we shall discuss  the entanglement properties of pure two mode Gaussian states. In particular the entanglement criterion and 
entanglement of formation for pure two mode Gaussian states will be considered. (Earlier works and references can be found for example in \cite{yk}.) For the special case of two mode Gaussian state the separability
criteria is both necessary and sufficient condition for entanglement. In order to obtain the entanglement criteria and entanglement of formation, 
the coefficients of the quadrature components of the wave function will be used \cite{rr}.

A two-mode Gaussian state is said to be entangled if we have a non-trivial cross term in the exponential, as for example :
 
\begin{equation}
 \psi = N_0 \exp{\left[-\frac{1}{2}\left(\alpha x_1^2+\beta x_2^2+2\gamma x_1x_2\right)\right]}, \ \ \ \  \gamma \ne0,
\label{2}
\end{equation}
where $\alpha$,$\beta$ and $\gamma$-s are in general complex numbers. Let us write them as $\alpha=\alpha_1+i\alpha_2$,
$\beta=\beta_1+i\beta_2$, $\gamma=\gamma_1+i\gamma_2$. From the normalization of these states we also have $\abs{N_0}^2=\Delta/\pi$ where $\Delta^2=\alpha_1\beta_1-\gamma_1^2>0$ (For details, see appendix A). In addition, we also have $\alpha_1>0$ to ensure that $\psi \in L^2$.
In  Ref. \cite{rr} the authors have demonstrated that if the vacuum state consists of a non-trivial cross term in the exponential (such as $\gamma x_1x_2$ in (\ref{2}) above), the state will be entangled. Consider the state in Eq. (\ref{2}), we define:

\bea
    A &=& 
\begin{pmatrix}
    \langle x_1^2 \rangle & \langle \anticommutator{x_1}{p_1}\rangle\\
    \langle \anticommutator{x_1}{p_1}\rangle & \langle p_1^2 \rangle
    \end{pmatrix},~~
    B = 
    \begin{pmatrix}
    \langle x_2^2 \rangle & \langle \anticommutator{x_2}{p_2}\rangle\\
    \langle \anticommutator{x_2}{p_2}\rangle & \langle p_2^2 \rangle
    \end{pmatrix}\label{19},\\
     C &=& 
    \begin{pmatrix}
    \langle x_1x_2 \rangle & \langle x_1p_2 \rangle\\
    \langle p_1x_2 \rangle & \langle p_1p_2 \rangle
    \end{pmatrix},~~
    J = 
    \begin{pmatrix}
    0 & 1\\
    -1 & 0
    \end{pmatrix}.
    \label{20}
\eea
The elements of the matrices are calculated for the Gaussian state in Eq. (\ref{2}). (Details are given in  Appendix A).
Let us define the quantity \cite{rr}:
\begin{equation}
    E_S=\det A \det B+(\frac{1}{4}-\abs{\det C})^2- Tr - \frac{1}{4}(\det A+\det B)
\end{equation}
where $Tr = tr(AJCJBJC^TJ)$. With this, Simon's seperability criterion \cite{rs} is stated as,
\begin{equation}
    E_S \geq 0.
\label{ec}
\end{equation}
This is a necessary and sufficient condition for bipartite Gaussian states to be separable; otherwise for $E_S<0$, the state is  entangled.
For the state in Eq. (\ref{2}), $E_S$ is calculated to be (using the results in Appendix A) \cite{rr}:

\begin{equation}
    E_S=-\frac{1}{4}\frac{\gamma_1^2+\gamma_2^2}{\Delta^2}
\label{es}
\end{equation}
where  $\Delta^2 = \alpha_1\beta_1 - \gamma_1^2>0$.  We can immediately see that for all non-zero $\gamma$, $E_S<0$ implying that 
the state will be entangled.

To quantify this entanglement, we pick an entropy measure which reduces to Von-Neumann entropy for a pure state: The Entanglement of Formation $E_F$. The mathematical definition and its physical interpretation has been discussed in the Appendix A . Here, we only give its form which is relevant for the state in Eq. (\ref{2}) \cite {rr}:

\begin{equation}
    E_F = (\Omega+1/2)ln(\Omega+1/2)-(\Omega-1/2)ln(\Omega-1/2)
\label{ef}
\end{equation}
where $\Omega$ is defined by the relation $\Omega^2=1/4-E_S$. We see that  for $E_S=0$ i.e. for the separable case $E_F$ is also zero as is expected.

In this paper, we are primarily interested in the induced Gaussian entanglement produced by the noncommutativity of the space. We have considered 
a particular anisotropic HO in two dimensional NC space.
We shall discuss about how the entanglement depends on the degree of anisotropy. The entanglement of the system is also quantified in terms of the deformation parameter $\theta$ and it has been shown that the entanglement vanishes if the deformation of the space is trivial. Moreover, we restate the separability criterion of the ground state of such a system. We will be using the results shown above.

\section{Anisotropic harmonic oscillator in two dimensional NC space}
 
In two dimension the NC extension of the quantum mechanics may be realized by a simple modification 
of the commutation relation among the self-adjoin position ($X_i, i=1,2$) and momentum ($P_i, i=1,2$) operators,
such that they satisfy the following commutation relations: 

\bea
[X_i, X_j]=i\epsilon_{ij}\theta,\ \ \ [X_i, P_j]=i \delta_{ij}, \ \ \ \ [P_i,P_j]=0, \ \ \ i,j=1,2,
\label{eha}
\eea  
with anti symmetric matrix  $\epsilon_{ij},~ \epsilon_{12}=1$.  
In this noncommutating space we consider an anisotropic oscillator represented by the following Hamiltonian:

\bea
H=\frac{P^2_1}{2m_1}+\frac{P^2_2}{2m_2}+\alpha_1X_1^2+\alpha_2X_2^2,\ \ \ \ \ \alpha_1 \ne \alpha_2,\ \  \alpha_1, \alpha_2 >0.
\label{H}
\eea
The purpose of the present
section is to find exact solution for this system in the two dimensional NC space. It should be noted that the solution of 
isotropic HO in arbitrary dimension in NC space has been considered previously in the literature \cite{id}.
The interesting property of the extended Heisenberg algebra of the type considered in Eq. (\ref{eha}) is that through a
linear transformation, popularly known as  Bopp's shift, it can be related to the standard Heisenberg algebra: 

\bea
[x_i, x_j]=0,\ \ \ [x_i, p_j]=i \delta_{ij}, \ \ \ \ [p_i,p_j]=0, \ \ \ i,j=1,2. 
\eea 
For our present purpose we consider the following Bopp's shift:

\bea
X_1=x_1-\frac{\theta}{2}p_2,\ \ 
X_2=x_2+\frac{\theta}{2}p_1,\ \  P_1=p_1,\ \ P_2=p_2.
\label{tr}
\eea
Obviously the transformation of Eq. (\ref{tr}) is not unitary as it changes the symplectic structure.
This transformation enables us to convert the Hamiltonian in the NC space into a modified Hamiltonian in the commutative space
having an explicit dependence on the deformation parameter $\theta$. The states of the system are then wave functions on the ordinary
Hilbert space. The dynamics of the system is now governed by the Schrodinger equation with the NC parameter $\theta$-dependent Hamiltonian. 

In the
present case, we cast the the Bopp shifted  Hamiltonian in the following form,

\bea
H= \frac{p_1^2}{2M_1}+\frac{p_2^2}{2M_2}+\frac{1}{2}M_1\omega_1^2x_1^2+\frac{1}{2}M_2\omega_2^2x_2^2
-\theta \left(\alpha_1x_1 p_2-\alpha_2x_2p_1\right),
\label{h}
\eea
where 

\bea
\frac{1}{M_1}=\left(\frac{1}{m_1}+\frac{\alpha_2\theta^2}{2}\right),\  \frac{1}{M_2}=\left(\frac{1}{m_2}+\frac{\alpha_1\theta^2}{2}\right);\ 
\frac{1}{2}M_i\omega_i^2=\alpha_i,\  i=1,2.
\label{map}
\eea
Let us make an important observation in connection to \cite{bl}. If the HO was isotropic, (as considered in \cite{bl}), then in (\ref{h}) the NC induced $\theta$-term is actually $\theta L_3$, that is angular momentum component in the $3$-direction. Since $L_3$ commutes with the isotropic HO, the $\theta L_3$-term will not make any changes in the isotropic HO wavefunctions and so, there will be no GE for isotropic HO in presence of noncommutativity of the space. We will show later that the explicit result for entanglement also corroborate this. However this result is in contradiction to  \cite{bl} where GE appears even for NC isotropic HO.

It is noteworthy that the introduction of noncommutativity modifies the mass parameters \cite{khp}. In general, the form of the Hamiltonian of Eq. (\ref{h}) depends on the map (\ref{tr}) \cite{cb}. It should be pointed out that modifications in  mass
parameters and its effect on the equivalence principle was considered previously in \cite{khp}. Furthermore it is  important to note that  although the Hamiltonian depends on the particular
map chosen,  the physical predictions  involving eigenvalues, expectation values, probabilities  do
not depend on the map, as proved in \cite{n4}. The Hamiltonian of Eq. (\ref{h}) can also be written in the following form:
\vskip .5cm
\bea
H&=&\frac{1}{2M_1}\left(p_1+\theta M_1\alpha_2x_2\right)^2+\frac{1}{2M_2}\left(p_2-\theta M_2\alpha_1x_1\right)^2\nonumber \\ 
&+& \frac{1}{2}\left(M_1\omega_1^2-M_2\theta^2\alpha_1^2\right)x_1^2+ \frac{1}{2}\left(M_2\omega_2^2-M_1\theta^2\alpha_2^2\right)x_2^2.
\label{hm}
\eea
For $m_1=m_2,~\alpha_1=\alpha_2 $, i.e for an isotropic HO in the NC space  (see Eq. (\ref{h})), this Hamiltonian has the interpretation of a particle
moving in a two dimensional plane (x,y) in a perpendicular magnetic field with a strength proportional to the noncommutating parameter $\theta$.
It is interesting to note that in this case, the effect of taking $\theta= 0$, is equivalent to switching off the magnetic field.\\

 In order to solve the Hamiltonian of Eq. (\ref{h}), it is convenient to express it in the following form \cite{ql}:

\bea
H=\frac{1}{2}X^T{\cal{H}}X, \ \ \ X=\left(x_1, p_1, x_2, p_2\right)^T,
\label{ham}
\eea
where the column matrix $X$ satisfies the commutator relations 

\bea
\left[X_{\alpha}, X_{\beta}\right]=-\left(\Sigma_y\right)_{\alpha \beta}
\label{sym}
\eea
 where $\Sigma_y=$diag$(\sigma_y, \sigma_y)$, $\sigma_y$ being the Pauli matrix, and ${\cal H}$ is the following symmetric matrix

\bea
{\cal H}=
\bp 
M_1\omega_1^2 & 0 & 0 & -\theta\alpha_1\\
0 & \frac{1}{M_1}& \theta \alpha_2 &0\\
0 & \theta \alpha_2 & M_2 \omega_2^2 & 0\\
 -\theta\alpha_1&0&0&  \frac{1}{M_2
 }
\ep.
\eea 
The idea is to diagonalized the Hamiltonian $H$ of Eq. (\ref{h}) such that it can be expressed as a sum of two one dimensional harmonic oscillators.
This diagonalization must be done  without affecting the symplectic structure of the system reflected in the Eq. (\ref{sym}), i.e. the linear transformation
that diagonalized ${\cal H}$ must preserve the relation (\ref{sym}). In the following procedure, we shall express the Hamiltonian into a form expressed in terms
of raising and lowering operators \cite{ql}. 
It should be noted here that the Hamiltonian $H$ and $X$ satisfy the following commutation relation

\bea
\left[iH, X\right]= \Omega X
\label{hx}
\eea 
where 
\bea
\Omega =i \Sigma_y {\cal H}
\label{oh}
\eea
having the explicit form

\bea
\Omega=
\bp
0 & \frac{1}{M_1} & \theta \alpha_2 & 0\\
-M_1\omega_1^2 & 0 & 0 & \theta \alpha_1\\
-\theta \alpha_1 & 0 & 0 & \frac{1}{M_2}\\
0 & -\theta \alpha_2 & -M_2\omega_2^2& 0
\ep.
\label{Om}
\eea
It should be noted that for $\theta \alpha_1=\theta \alpha_2=\omega_{B}$ and $M_1=M_2$, the matrix $\Omega$ reduces to the form considered in the Ref. \cite{ql}
for the case of isotropic oscillator. Now the next step is to diagonalized the matrix $\Omega$. Rest of the calculations are quite involved and we relegate these to Appendix B.

The diagonal Hamiltonian is given by,
\bea
H=\sigma_1\left(a_1^{\dagger}a_1+\frac{1}{2}\right)+ \sigma_2\left(a_2^{\dagger}a_2+\frac{1}{2}\right),
\label{dha}
\eea
where
\bea
\sigma_1=\left(\frac{b+\sqrt{D}}{2}\right)^{\frac{1}{2}}, \ \ \  \sigma_2=\left(\frac{b-\sqrt{D}}{2}\right)^{\frac{1}{2}},\ \ \ 
\eea
with $\sigma_1> \sigma_2$. Explicit expressions for $b,D$ are given in Appendix B.

Clearly the Hamiltonian of Eq. (\ref{h}) is expressed as a sum of two decoupled one dimensional 
HOs. This Hamiltonian has the same form as considered in Ref. \cite{ql}. However, in this case the 
expressions for $\sigma_1$ and $\sigma_2$ are different which incorporate the effect of anisotropic nature of the system. 
The energy levels are readily obtained as

\bea
E_{n_1,n_2}=\sigma_1\left(n_1+\frac{1}{2}\right)+\sigma_2\left(n_2+\frac{1}{2}\right),\ \ \ \ n_1,n_2=0,1,2,3,.....
\label{en}
\eea 
which is the sum of the energy of two one dimensional harmonic oscillators. 

The energy eigenfunctions  for the system in coordinate representation are straightforward to obtain. In particular the ground state is given by
\bea
\psi_{00}=N_0 \exp{\left[-\frac{1}{2}\left(\Lambda_{11}x_1^2+\Lambda_{22}x_2^2+(\Lambda_{12}+\Lambda_{21})x_1x_2\right)\right]}.
\label{gs}
\eea
The remaining task is is to compute the entanglement. 
Explicit expressions of $\Lambda_{ij}$ and some steps of computation are provided in Appendix B.

\section {Entanglement for the anisotropic oscillator in NC space}

After obtaining the exact solution for the ground state wave function of the anisotropic oscillator in NC space, we are now in the position to study 
the entanglement properties of the system. As we have already emphasized in Section 2. Eq. (\ref{es}), the $x_1x_2$-term in Eq. (\ref{gs}) is a signature of entanglement.

In order to quantify the entanglement of the system we shall calculate the quantity $E_S$, described in
section 2 (see Eq. (\ref{es})), which takes the following form in the present case:
\begin{equation}
    E_S=-\frac{1}{4}\frac{[\operatorname{Im}(\Lambda_{12})]^2}{(\Lambda_{11}\Lambda_{22})},
\label{es1}
\end{equation}
with the Simon's Separability criterion\cite{rs}:

\begin{equation}
    E_S \geq 0.
\label{26}
\end{equation}
From the expression of $\Lambda_{12}$ given in Appendix B, we see that $\Lambda_{12}$ vanishes for $\theta=0$ showing that GE is induced by noncommutativity of the space. One of the main feature is that this entanglement also depends on the degree of the anisotropy. Furthermore, the important point is that for $m_1=m_2$
and $\alpha_1=\alpha_2$, the off diagonal terms $\Lambda_{ij}, i\ne j$ vanish which again shows that for isotropic case there is no entanglement.

Therefore, the necessary and sufficient condition for entanglement is to have a negative $E_S$\cite{oe}\cite{ar}. In our present case, 
Eq. (\ref{es1}) takes the following form:

\begin{equation}\label{ess}
    E_S= -\frac{\theta^2}{8} \sqrt{\alpha_1m_2\alpha_2 m_1}\frac{(\sqrt{\alpha_1 m_2}-\sqrt{\alpha_2 m_1})^2}{[2\theta^2 \alpha_1 m_2\alpha_2 m_1+(\sqrt{\alpha_1 m_2}+\sqrt{\alpha_2 m_1})^2]}.
\end{equation}
It is evident from the above expression that $E_S$ can never be positive. However, it can be zero. This situation arises when either $\theta=0$ or $\frac{\alpha_1}{m_1}-\frac{\alpha_2}{m_2}=0$.
For $\theta=0$ i.e. in the absence of deformation, the system of harmonic oscillators is defined on the usual commutative space and as expected there is no entanglement in this case:
\begin{equation}
    \psi_{00}(x_1,x_2) \sim exp[-1/2(\Lambda_{11}x_1^2+\Lambda_{22}x_2^2)]
    =exp[-1/2\Lambda_{11}x_1^2]\otimes exp[-1/2\Lambda_{22}x_2^2].\nonumber
\end{equation}
Furthermore, $E_s$ can be zero for NC anisotropic HO provided the condition $ \frac{\alpha_1}{m_1}=\frac{\alpha_2}{m_2}$ is satisfied. Again we reaffirm that $E_s$ vanishes for isotropic HO even for non-zero $\theta$.

We can consider two special cases:\\
(i) $\alpha_1=\alpha_2=\alpha$. In this case 
\begin{equation}\label{n1}
E_S^{(i)}= -\frac{\theta^2\alpha m_1m_2}{8{\sqrt{m_1m_2}}} 
	\frac{(\sqrt{m_1}-\sqrt{m_2})^2}{[2\theta^2\alpha m_1m_2+(\sqrt{m_1}+\sqrt{m_2})^2]}.
\end{equation}
(ii) $m_1=m_2=m$. In this case 
\begin{equation}\label{n2}
E_S^{(ii)}= -\frac{\theta^2m\alpha_1\alpha_2}{8{\sqrt{\alpha_1\alpha_2}}} 
\frac{(\sqrt{\alpha_1}-\sqrt{\alpha_2})^2}{[2\theta^2m\alpha_1\alpha_2 +(\sqrt{\alpha_1}+\sqrt{\alpha_2})^2]}.
\end{equation}
Note that case (i) can be identified with a system of anisotropic oscillator ($m_1\ne m_2$) placed in a magnetic field, perpendicular to the oscillator plane. On the other hand, case (ii) corresponds to a conventional anisotropic oscillator ($\alpha_1\ne\alpha_2$). The former case (i) might be amenable to experimental verification. Also it is interesting to note the symmetry between (\ref{n1}) and (\ref{n2}) under the interchange of mass and frequency parameters.

For the non-trivial deformation case, the separability criterion, therefore, boils down to 
\begin{equation}
    \frac{\alpha_1}{m_1}=\frac{\alpha_2}{m_2}.
\end{equation}
We can crosscheck the validity of this criterion for isotropic harmonic oscillators with equal masses. The above mentioned system will be separable as the Hamiltonian for this case can be written as:
\begin{equation}
    H=H_0-\theta\alpha L_3
\end{equation}
where, 
\begin{equation}
    H_0=\frac{1}{2}(\frac{1}{m}+\frac{\alpha\theta^2}{2})(p_1^2+p_2^2)+\alpha(x_1^2+x_2^2)
\end{equation}
and,
\begin{equation}
    L_3=x_1p_2-x_2p_1.
\end{equation}
As $L_3$ commutes with $H_0$ hence H shares same eigenstates with $H_0$ which are separable. This can be easily verified from our separability criterion.

Let us now  plot the variation of $E_{F}$ of Eq. (\ref{ef}) with increasing deformation $\theta$ and increasing ratio $r=\frac{\alpha_1/m_1}{\alpha_2/m_2}$. Note that $r$ can be considered as a generalized measure of anisotropy, that is $r=1$ trivially for  isotropic case $m_1=m_2,~\alpha_1=\alpha_2$. However, $r$ can be unity for anisotropic case as well provided $\alpha_1/m_1= \alpha_2/m_2$. Fig.1 and Fig.2 describe the variation of the function $E_F$ with $\theta$ and $r$ respectively. 
\begin{figure}
\centering
\parbox{5cm}{
\includegraphics[width=5cm]{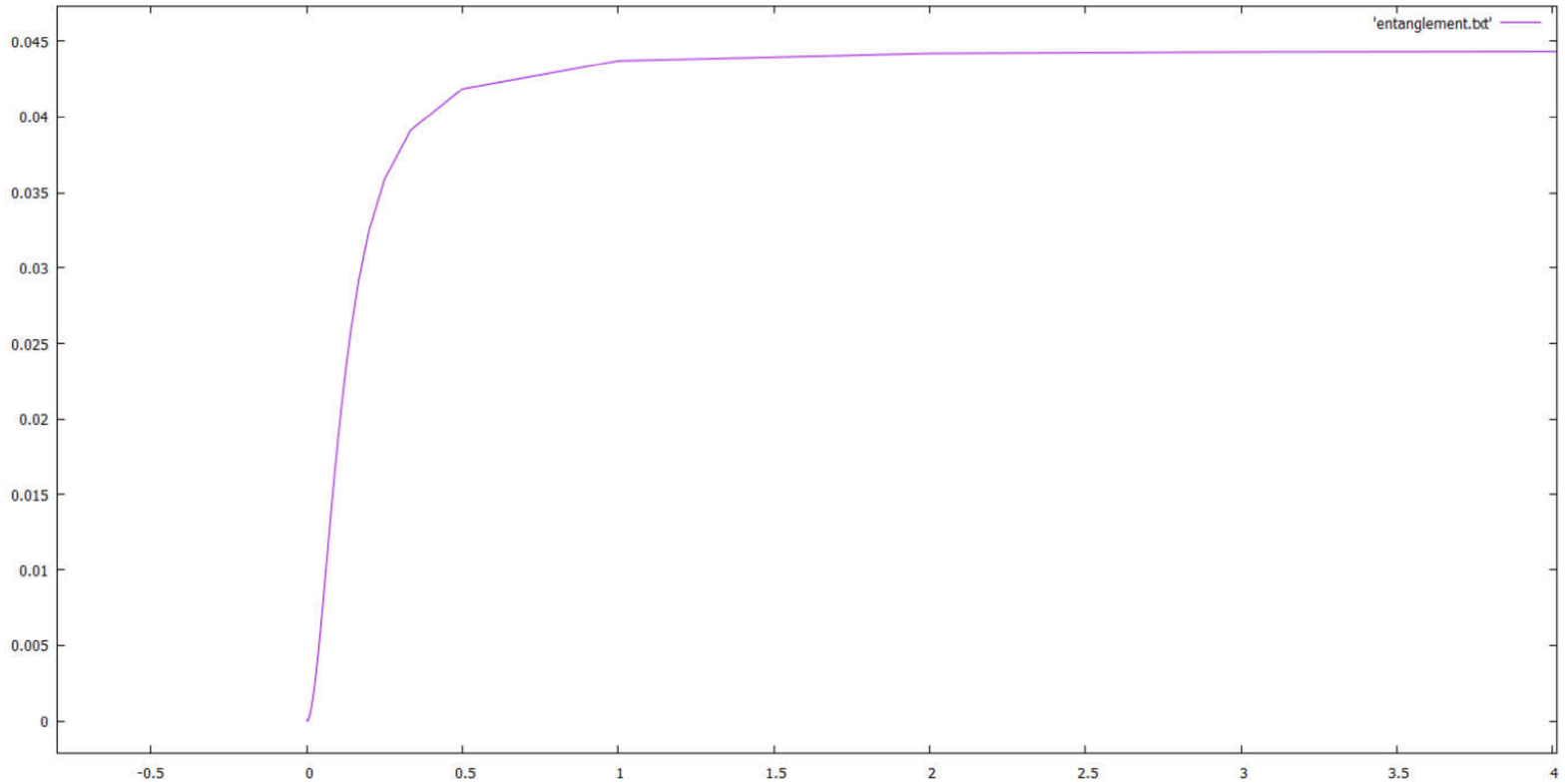}
\caption{ $E_F$ (ordinate) vs $\theta$ (abcissa) for $m_1=m_2=1$, $\alpha_1=5$ and $\alpha_2=10$ }
\label{fig1figsA}}
\qquad
\begin{minipage}{4cm}
\includegraphics[width=5cm]{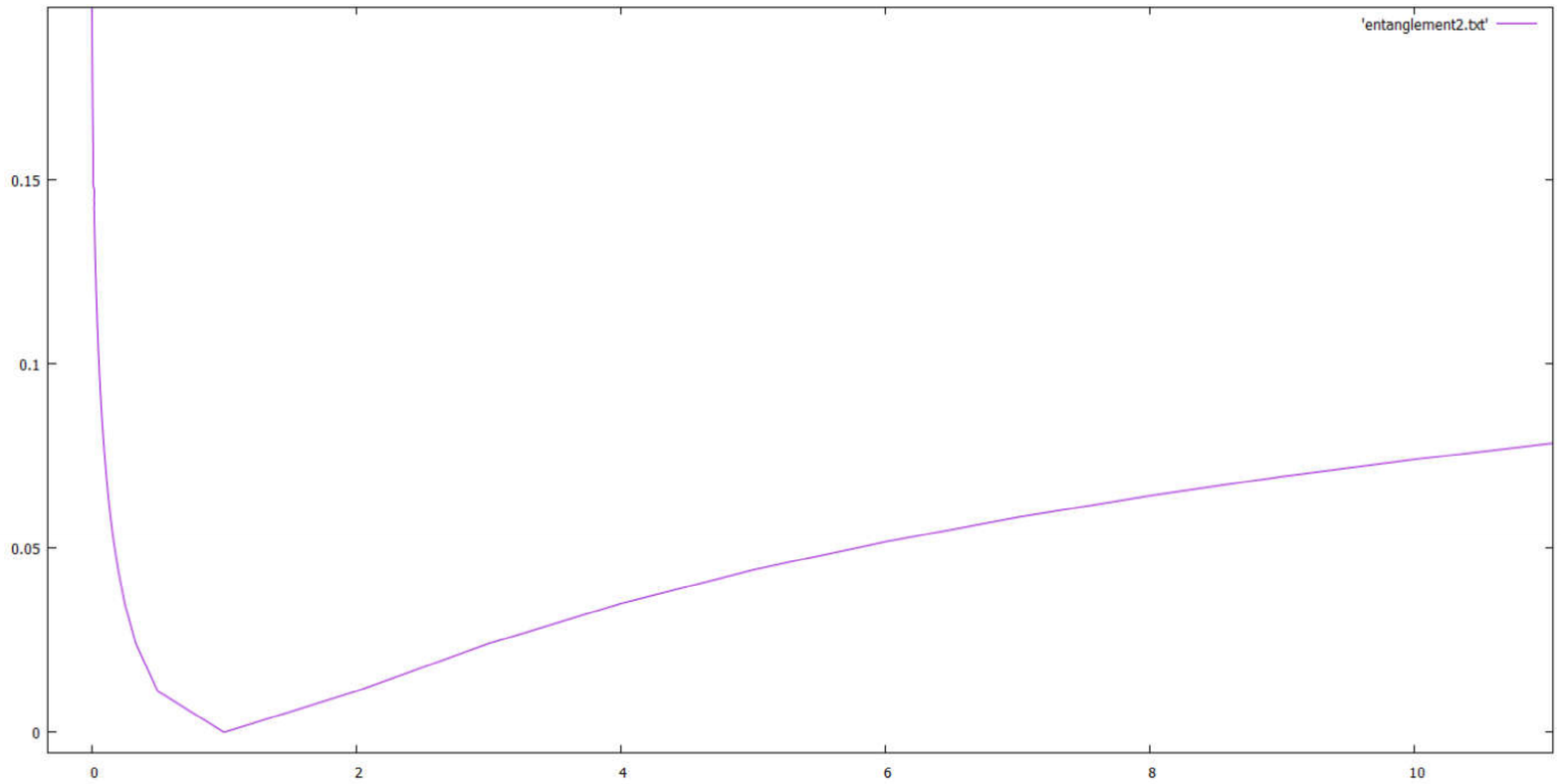}
\caption{$E_F$ (ordinate) vs $r$ (abcissa) for $m_1=m_2=1$, $\theta=1$ and $\alpha_1\alpha_2=2$}
\label{fig1figsB}
\end{minipage}
\end{figure}
For the plot $\theta$ vs $E_F$, we have taken unit mass for both oscillators and $\alpha_1=5$, $\alpha_2=10$, i.e., $r=0.5$. In the 2nd figure we have plotted r vs $E_S$ where, we have fixed $m_1$, $m_2$ and $\theta$ to unity and an additional criterion of $\alpha_1\alpha_2=2$ without loss of generality.

We see that the function increases rapidly for lower values of $\theta$, and saturates asymptotically for high $\theta$. The bound is derived straightforwardly by taking the limit of $\theta$ going to infinity in (\ref{ess}) for $E_S$:
\begin{eqnarray}
    \lim_{\theta\to\infty}E_S(\theta)=-\frac{1}{16}\frac{1}{\sqrt{\alpha_1m_2\alpha_2m_1}}(\sqrt{\alpha_1m_2}-\sqrt{\alpha_2m_1})^2.
\end{eqnarray}
Now as $\Omega$ is related to $E_S$ by the equation $\Omega^2=1/4-E_S$, hence the bound on $\Omega$ can be obtained by simply putting the limiting value of $E_S$ in the equation:
\begin{align}\label{omega}
    \lim_{\theta\to\infty}\Omega= \Omega_0 =    \frac{1}{4}(\alpha_1m_2\alpha_2m_1)^{-1/4}(\sqrt{\alpha_1m_2}+\sqrt{\alpha_2m_1}).
\end{align}
Hence, the bound on $E_F$ is given by:
\begin{equation}
    E_F < (\Omega_0+1/2)ln(\Omega_0+1/2)-(\Omega_0-1/2)ln(\Omega_0-1/2).
\end{equation}
On the other hand, $r$ being the measure of anisotropy, it is important to find the profile of $E_S$ against $r$. The plot of $r$ versus $E_F$ has two regions $r<1$ and $r>1$. As we converge to the point $r=1$, we see that it vanishes which is exactly what we prescribe as the separability criterion, but away from the point $r=1$, where the degree of anisotropy is non-trivial, the entanglement increases rapidly. It is clear from (\ref{omega}) that  the measure of the entanglement is symmetric to the interchange of the oscillators.  
 
\section{Conclusion}

In this paper we have studied in detail the entanglement property of a two dimensional  anisotropic harmonic oscillator, where the entanglement is induced by spatial noncommutativity. We follow the procedure of Bopp's shift, a transformation that reduces the noncommutative model to one comprising of canonical variables. An additional interaction term, dependent up on the noncommutativity parameter $\theta$, is generated. This all important term is responsible for the entanglement. The essential role of anisotropy is emphasized throughout the paper and we have also pointed out the mismatch between our results and that of \cite{bl}. Since the two computational schemes of \cite{bl} and the present one are inherently different, the choice between the two procedures can only be settled through experiments.

As an intermediate step we have obtained exact solutions for the energy eigenvalues and eigenfunctions of a planar anisotropic oscillator, which is a new result in the context of NC geometry induced physics. The nontrivial entanglement is established by  considering Simon's separability criterion. We have shown dependence of Entanglement of Fomation upon the noncommutative deformation $\theta$ and also upon the anisotropy parameter $r$. We have clearly established that the entanglement is zero for vanishing noncommutative deformation, $\theta =0$. However,  anisotropy plays an essential role for generating entanglement since in noncommutative deformed case for an isotropic oscillator, the eigenfunctions of the deformed and undeformed oscillator are the same which shows the there can be no entanglement for an isotropic oscillator, even for nonzero noncommutativity.

Finally, we briefly discuss  areas where  the results, derived in this paper, can be tested experimentally. The anisotropic oscillator model has a wide range of applications. For instance, it can describe motion of an electron in an anisotropic metal lattice \cite{ncc}. In the presence of  external electromagnetic fields,  these model might play a role in semiconductor physics. Diamagnetic properties of small metallic particles are mapped to these models \cite{nc3}. An empirical polarizable force field, based on the classical Drude
oscillator model, is another area of application \cite{nc4}. Furthermore, in the recently developed theory of magneto-optical phenomena, anisotropic oscillator models have played significant roles  \cite{nc2}. Probing entanglement properties in these experimentally realizable systems can give new insights.

Let us conclude our work with a cautionary note with a mention of certain features of our framework that can be improved. Note that we have performed our analysis in terms of canonical variables $x,p$ obeying canonical commutation relations, even though the original model was posited in terms of $X,P$ degrees of freedom that satisfy the non-canonical (or non-commutative) commutation relations. Also, in this context, recall that the structure of the noncommutative Hamiltonian is identical to the canonical one, with the variables $x,p$ replaced by $X,P$. Indeed, in all instances \cite{n8, n11, n16, n3} of noncommutative extension of quantum mechanics in perturbative framework this starting point is an assumption since we do not really know of any typical dynamical model explicitly in noncommutative space. Recall that in a similar vein, in noncommutative generalization of a quantum field theory as well the basic model is that of a known canonical field theory model with products of field variables (at a same spacetime point) replaced by the so called $*$-product (or Groenwald-Moller product) \cite{sw, ncrev, n9, dv,s1,s2,s3,s4,s5}. For want of a better option, in all examples of noncommutative quantum mechanical or field theory models, the next step is to express the noncommutative models in terms of canonical degrees of freedom, through Darboux map in quantum mechanics (as done in here and in other works) or through canonical fields in a field theory by expanding the $*$-product (and incorporating the Seiberg-Witten map for gauge theories  \cite{sw, ncrev, s1}. This procedure allows us to exploit standard computational techniques on the Hamiltonian (in quantum mechanics) or Lagrangian (in quantum field theory) that receive additional terms induced by noncommutativity through the Darboux map (in quantum mechanics) or $*$-product (in quantum field theory). Thus the final results are generically computed in a canonical framework in terms of canonical variables. Precisely this scheme has been followed in our work. A notable work, relevant to the present paper, is by Bastos et. al. in \cite{n6}, where also Darboux map is used to convert the noncommutative model to an extended model in terms of canonical variables in a Wigner function approach but importantly, it is shown that the final result can be reexpressed in terms of noncommutative variables and also the final results are independent of the Darboux map. However, we have done our analysis in a different framework, and indeed,  it would  be nice if in our case also similar procedure can be adapted. This is an interesting open question that we wish to pursue in near future.
	
	Another point is the dependency of our results on the specific Darboux map. We have exploited the simplest and most direct form of map that has been used by others in numerous occasions. We admit that one can use more elaborate maps that can change the numerical results but it is quite obvious that the take home message of our paper, that entanglement can be generated through noncommutativity, will remain intact. Only experimental verification can fix a specific choice of the  map. Furthermore, apart from numerical difference in the final result of entanglement measure, more complicated Darboux maps can introduce qualitative   changes in the final result, but it is expected that those will involve corrections involving higher orders of the noncommutative parameter $\theta$. So far our results are exact with the entanglement measure appearing in  (\ref{ess}) in $O(\theta^2)$ and corrections to still higher orders in $\theta$ may not be significant for small $\theta$ in a perturbative framework.
\vskip .3cm

{\bf{Acknowledgement:}} We thank the Reviewers for  their insightful comments that have helped us to improve the paper in a significant way.

\vskip .3cm

\section{Appendix A}

Normalization of the state in Eq.(\ref{2}) produces the criterion that $\alpha_1\beta_1-\gamma_1^2$ and $\alpha_1$ must be positive:
\begin{equation} 
\begin{split}
|\psi|^2 & = |N_0|^2\int exp[-(\alpha_1q_1^2+\beta_1q_2^2+2\gamma_1q_1q_2)] dq_1 dq_2 \\
 & = |N_0|^2\int\sqrt{\frac{\pi}{\alpha_1}}exp[-(\beta_1-\frac{\gamma_1^2}{\alpha_1})q_2^2] dq_2 \\
 & = |N_0|^2 \sqrt{\frac{\pi}{\alpha_1}} \sqrt{\frac{\pi}{\beta_1-\frac{\gamma_1^2}{\alpha_1}}} \\
 & = |N_0|^2 \frac{\pi}{\sqrt{\alpha_1\beta_1-\gamma_1^2}}.
\end{split}
\end{equation}
This should equal to unity. Hence, we have $|N_0|^2=\frac{\sqrt{\alpha_1\beta_1-\gamma_1^2}}{\pi}=\frac{\Delta}{\pi}$. As the Gaussian Integral must be convergent, therefore $\alpha_1\beta_1-\gamma_1^2$ must be greater than zero. So, we have a real $\abs{N_0}^2$.

Now, the matrices A, B and C are defined in Eq.(\ref{19}) and in Eq.(\ref{20}) and the entries are calculated for the state in (\ref{2})\cite{rr}:
\begin{equation}
    \langle x_1^2 \rangle=\frac{\beta_1}{2\Delta^2};\ \ \ \
    \langle x_2^2 \rangle=\frac{\alpha_1}{2\Delta^2};\ \ \ \
    \langle x_1x_2 \rangle=-\frac{\gamma_1}{2\Delta^2};\ \ \ \
\end{equation}
\begin{align}
    \langle p_1^2 \rangle &=\frac{\beta_1\abs{\alpha}^2-\alpha_1(\gamma_1^2-\gamma_2^2)-2\alpha_2\gamma_1\gamma_2}{2\Delta^2};\\
    \langle p_2^2 \rangle &=\frac{\alpha_1\abs{\beta}^2-\beta_1(\gamma_1^2-\gamma_2^2)-2\beta_2\gamma_1\gamma_2}{2\Delta^2};\\
    \langle p_1p_2 \rangle &=\frac{(\alpha_1\gamma_1+\alpha_2\gamma_2)\Delta^2+(\alpha_1\beta_2-\gamma_1\gamma_2)(\alpha_1\gamma_2-\alpha_2\gamma_1)}{2\alpha_1\Delta^2}
\end{align}
\begin{align}
    \langle \anticommutator{x_1}{p_1}\rangle &=\frac{\gamma_1\gamma_2-\alpha_2\beta_1}{2\Delta^2};
    \langle \anticommutator{x_2}{p_2}\rangle =\frac{\gamma_1\gamma_2-\alpha_1\beta_2}{2\Delta^2};\\
    \langle x_1p_2 \rangle &=\frac{\gamma_1\beta_2-\gamma_2\beta_1}{2\Delta^2};
    \langle x_2p_1 \rangle =\frac{\gamma_1\alpha_2-\gamma_2\alpha_1}{2\Delta^2}.
\end{align} 
From these relations we can calculate the determinants of A, B and C in terms of the quadrature coefficients:
\begin{align}
    \det A = \langle x_1^2 \rangle \langle p_1^2 \rangle - \langle \anticommutator{x_1}{p_1}\rangle^2  \\
    = \det B = \langle x_2^2 \rangle \langle p_2^2 \rangle - \langle \anticommutator{x_2}{p_2}\rangle^2 \\
    = \frac{1}{4}+\frac{1}{4}\frac{\gamma_1^2+\gamma_2^2}{\Delta^2}\label{a7}
\end{align}
\begin{equation}
    \det C = -\frac{1}{4}\frac{\gamma_1^2+\gamma_2^2}{\Delta^2}.
\end{equation}

\vspace{0.3in}

In this paper, we quantified the entanglement using the Entanglement of Formation. If we have $n$ number of Bell states and if by local operations and classical communication (LOCC) we can produce only $m$ number of the given state $\ket{\psi}$ from those Bell states then the entanglement of formation corresponds to the ratio n/m\cite{nc}. Mathematically it can be defined as\cite{rr}:

\begin{equation}\label{23}
E_F(\rho)=inf[\sum_K p_K E_{vn}(\psi_{K})]
\end{equation}
for all possible decomposition $\rho=\sum_K p_K \ket{\psi_K}\bra{\psi_K}$. $E_{vn}$ refers to the Von-Neumann entropy of the state which can 
be considered as the quantum mechanical analogue of Shannon entropy. Shannon entropy of X quantifies the amount of uncertainty about X before 
we learn its value\cite{nc}. For an entangled state, we will always have a decreased entropy as a measurement of one of its subsystem can give 
information about the other subsystem therefore decreasing its uncertainty\cite{nc}.
However, this formula is rarely used as there are difficulties in optimization involved in finding the infimum\cite{rr}. Fortunately, in this case, it can be derived in a simpler form which is mentioned in Eq.(\ref{ef}).

\section{Appendix B}

In this section we shall consider the diagonalization of the matrix $\Omega$ considered in Eq. (\ref{Om}) and will present the necessary steps to arrive
at the Eq. (\ref{dha}). To diagonalize $\Omega$, note that it 
is not a symmetric matrix. The characteristic polynomial for this matrix is given by

\bea
\det{\left(\lambda I-\Omega\right)}=\lambda^4 +b \lambda^2+c =0,
\eea 
where

\bea
b=\omega_1^2+\omega_2^2+2\theta^2\alpha_1\alpha_2,\ \  c= \left(\omega_2^2-\theta^2 \frac{M_1}{M_2} \alpha^2_2\right)
\left(\omega_1^2-\theta^2 \frac{M_2}{M_1} \alpha^2_1\right).
\eea
From Eq. (\ref{map}), it can easily be checked that $c>0$ 
and $D\equiv b^2-4c\ge0$ with $\sqrt{D}<b$. Therefore, both roots for $\lambda^2$ are real and negative,
implying that the above characteristic equation has four pure imaginary roots: 
\bea
\{\lambda_1, \lambda_2, \lambda_3, \lambda_4\}=\{-i\sigma_1,i\sigma_1, -i\sigma_2, i\sigma_2\}
\eea
where

\bea
\sigma_1=\left(\frac{b+\sqrt{D}}{2}\right)^{\frac{1}{2}}, \ \ \  \sigma_2=\left(\frac{b-\sqrt{D}}{2}\right)^{\frac{1}{2}},\ \ \ 
\eea
with $\sigma_1> \sigma_2$. 
Since the matrix $\Omega$ is not symmetric the left and right eigenvectors are different. The two left eigenvectors corresponding to the 
eigenvalues $-i\sigma_i (i=1, 2)$ are given by 
\bea
u_i\Omega=-i\sigma_i u_i,\ \ \ \ i=1, 2,
\label{eu}
\eea
and the other two are $u_i^*$ corresponding to the eigenvalues $i\sigma_i (i=1, 2)$. 
In the explicit form the left eigenvectors are 

\bea
u_i=\frac{1}{k_i} 
\bp
-i M_1M_2\sigma_i(\sigma_i^2-\omega_2^2-\theta^2\alpha_1\alpha_2)\\
M_2(\sigma_i^2-\omega_2^2)+\theta^2 M_1 \alpha_2^2\\
\theta M_1M_2 \alpha_2(\sigma_i^2-\theta^2\alpha_1\alpha_2)+\theta M_2^2\alpha_1 \omega_2^2\\
i\theta \sigma_i (M_1\alpha_2+M_2\alpha_1)
\ep^T,\ \ \ i=1, 2,
\eea
where $k_i$'s are the normalization constants. The right eigenvectors are $v_i, i=1,2$ (corresponding to the eigenvalues $-i\sigma_i, i=1, 2$) and 
$v^*_i, i=1,2$ (corresponding to the eigenvalues $-i\sigma_i, i=1, 2$)  satisfying the equation

\bea
\Omega v_i=-i\sigma_i v_i,\ \ \ \ i=1, 2.
\label{ev}
\eea
From Eqs. (\ref{eu}) and (\ref{ev}), it is easy to check that 

\bea
u^*_iv_j=u_iv_j^*,\ \forall \  i, j=1,2.
\eea
and by appropriately choosing the normalization constant, we have
\bea
u_iv_j=u^*_iv_j^*=\delta_{ij}, \forall \ i,j=1,2.
\eea
Using Eqs. (\ref{hx}) and (\ref{eu}, \ref{ev}), the relation between the left and right eigenvectors may be expressed as \cite{ql}
\bea
v_i= -\Sigma_y u^{\dagger}_i, \ \ i=1,2.
\eea
We define a $4\times 4$ row matrix

\bea
Q=\left(v_1, v_1^*, v_2, v_2^*\right).
\label{q}
\eea 
The inverse is given by 

\bea
Q^{-1}=\left(u_1^T, {u_1^*}^T, u_2^T, {u_2^*}^T\right)^T,
\label{qi}
\eea
and 

\bea
Q^{-1}\Omega Q=&diag& \left(-i\sigma_1, i\sigma_1,-i \sigma_2, i\sigma_2\right).
\eea
Another important relation is 
\bea
Q^{\dagger}=-\Sigma_z Q^{-1}\Sigma_y,\ \ \Sigma_z= &diag&{\left(\sigma_z, \sigma_z\right)}.
\label{qd}
\eea
Where, $\sigma_x$, $\sigma_y$ and $\sigma_z$ are the Pauli matrices.\\
We now define the raising and lowering operators by the relations

\bea
a_i=u_i X,\ \ a_i^{\dagger}=u_i^*X,\ \ \left[a_i, a_j^{\dagger}\right]=\delta_{ij}, \ \ i, j \ =1,2.
\eea
We define

\bea
A=\left(a_1, a_1^{\dagger}, a_2, a_2^{\dagger}\right)
\eea
which can be written as 

\bea
A=Q^{-1}X.
\label{ax}
\eea
Now using relations (\ref{ham}), (\ref{oh}) and (\ref{qd}), it is easy to show that

\bea
H=\frac{1}{2}A^{\dagger}\Sigma A,\ \ \ \ \Sigma=&diag&\left(\sigma_1, \sigma_1, \sigma_2,\sigma_2\right)
\eea
which written in the explicit form yields the Hamiltonian of Eq. (\ref{dha}).

Now, we shall work out the wave functions for the system described by the Hamiltonian of Eq. (\ref{h}). This may be expressed as:
\bea
\ket{n_1n_2}=\frac{1}{\sqrt{n_1!n_2!}}(a_1^\dagger)^{n_1}(a_2^\dagger)^{n_2}\ket{00},\ \ \ \ n_1,n_2=0,1,2,3,.....
\label{wf}
\eea
where the ground state $\ket{00}$ is given by

\bea
a_1\ket{00}=a_2\ket{00}=0.
\label{a1a2}
\eea
Next, we shall determine this ground state wave function $\psi_{00}$ in the coordinate space representation where

\bea
\psi_{00}=\braket{x_1x_2|n_1n_2}
\eea
and the other excited states can be obtained by using Eq. (\ref{wf}). In the coordinate space representation, Eq. (\ref{a1a2}) takes
the following form:

\bea
\left(\xi_{ij}x_j-i \eta_{ij}\partial_j\right)\psi_{00}=0
\label{gse}
\eea
where

\bea
\xi=
\bp
u_{11} & u_{13}\\
u_{21} &  u_{23}
\ep
,\ \ \ 
\eta=
\bp
u_{12}  &  u_{14}\\
u_{22}&  u_{24}
\ep
\eea
with $ u_{i\beta}(i=1,2; \beta=1,2,3,4)$ is the $\beta$-th component of $u_i$. In order to solve Eq. (\ref{gse}), we choose the trial solution
$\psi_{00}=N_0 \exp{(-S(x_1,x_2))}$, where $N_0$ is the normalization constant and $S=\frac{1}{2}x_i \Lambda_{ij}x_j$. Substituting this into Eq. (\ref{gse}), we get

\bea
\Lambda=i\eta^{-1}\xi.
\eea
This can easily be solved with the following expression for the components of the matrix $\Lambda$:

\bea
\Lambda_{11} &=& \frac{M_1M_2\sigma_1\sigma_2 (\sigma_1+\sigma_2)}{\left[M_2(\omega^2_2+\sigma_1\sigma_2)-\theta^2M_1\alpha_2^2\right]},\\
\Lambda_{22}&=& \frac{M_2(M_2\omega_2^2-M_1\theta^2\alpha_2^2) (\sigma_1+\sigma_2)}{\left[M_2(\omega^2_2+\sigma_1\sigma_2)-\theta^2M_1\alpha_2^2\right]},\\
\Lambda_{12}&=& \Lambda_{21} = \frac{iM_2(\theta^3M_1\alpha_2^2\alpha_1-\theta M_2\alpha_1\omega_2^2+\theta M_1\alpha_2\sigma_1\sigma_2)}{[M_2(\omega_2^2+\sigma_1\sigma_2)-\theta^2M_1\alpha_2^2]}. 
\eea
All these expressions reduce to the form as obtained in the Ref. \cite{ql} for the case $M_1=M_2$ and $\theta \alpha_1=\theta \alpha_2=\omega_{B}$.


\end{document}